\documentclass[wssquare]{ws-procs975x65}

\newcommand\neededonlyforarXiv[2]{#1}

\newcommand\prerefereechanges[1]{#1}

\newcommand\hMpc{\mbox{$h^{-1}$ Mpc}}

\newcommand\fvir{f_{\mathrm{vir}}}

\newcommand\Ommzero{\Omega_{\mathrm{m0}}}
\newcommand\OmLam{\Omega_{\Lambda}} 
\newcommand\OmLamzero{\Omega_{\Lambda0}} 

\newcommand\OmegakClarkson{\widehat{\Omega}_k}

\providecommand\lsim{\mathop{\hbox{${\lower3.8pt\hbox{$<$}}\atop{\raise0.2pt\hbox{$\sim$}}
$}}} 
\providecommand\gsim{\mathop{\hbox{${\lower3.8pt\hbox{$>$}}\atop{\raise0.2pt\hbox{$
          \sim$}}$}}}

\providecommand\colorrgb[1]{\color[rgb]{#1}} 
\providecommand\rotatebox[2]{#2}

\providecommand\hMpc{\mbox{$h^{-1}$ Mpc}}

\providecommand\jcap{Journ.~Cosm.~Astr.~Phys.}
\providecommand\apjs{Astrophys.~J.~Supp.}                 
\providecommand\apj{Astrophys.~J.}                 
\providecommand\prd{Phys.~Rev.~D}
\providecommand\mnras{Mon.~Not.~Roy.~Astr.~Soc.}
\providecommand\aap{Astron.~Astroph.}

\providecommand\cqg{Class.~Quant.~Gra.}   %
\providecommand\grg{General Relativity and Gravitation}

\providecommand\ijmpd{Int.~J.~Mod.~Phys.~D}

\newcommand{\CE}{{\cal E}}

\newcommand{\CM}{{\cal M}}

\providecommand\inlinecite[1]{Ref.~\citenum{#1}}

\neededonlyforarXiv{
  \providecommand\eprint[1]{\href{http://arXiv.org/abs/#1}{arXiv:#1}} 

  \usepackage{hyperref}
  \hypersetup{
    colorlinks=true,
    citecolor=blue,
    linkcolor=magenta,
    filecolor=cyan,      
    urlcolor=cyan,
  }
}{
  \providecommand\eprint[1]{{\tt [arXiv:#1]}}
  \providecommand\href[2]{\url{#1}} 
}

\usepackage[varg]{txfonts}
\usepackage{fixltx2e} 

\begin{document}

\neededonlyforarXiv{
}{
}

\markboth{Boudewijn F. Roukema}{BAO peak flexible standard ruler}

\title{The baryon acoustic oscillation peak: a flexible standard ruler}

\author{Boudewijn F. Roukema} 

\address{Toru\'n Centre for Astronomy, Faculty of Physics, Astronomy and Informatics, Grudziadzka 5,\\Nicolaus Copernicus University, ul. Gagarina 11, PL--87--100 Toru\'n, Poland \\
Universit\'e de Lyon, Observatoire de Lyon, 
Centre de Recherche Astrophysique de Lyon, \\CNRS UMR 5574: Universit\'e Lyon~1 and \'Ecole Normale Sup\'erieure de Lyon,\\
9 avenue Charles Andr\'e, F--69230 Saint--Genis--Laval, France\footnote{During visiting lectureship.}}


\begin{abstract}
  {For about a decade,} 
  the baryon acoustic oscillation (BAO) peak at about 105{\hMpc} has
  provided a standard ruler test of the $\Lambda$CDM 
  {cosmological} model,
  a member of the Friedmann--Lema\^{\i}tre--Robertson--Walker (FLRW) family of
  cosmological models---according
  to which comoving space is rigid.
  However, general relativity does not require
  comoving space to be rigid. During the virialisation epoch,
  {when} 
  the most massive structures form by gravitational collapse, it
  should be expected that comoving space evolves inhomogeneous curvature as structure
  grows.  The BAO peak standard ruler should also follow this
  inhomogeneous evolution if the comoving rigidity assumption is false. This 
  ``standard'' ruler has now been detected to be flexible, as expected under
  general relativity.
\end{abstract}


\bodymatter


\section{Structure formation should curve space}

For about a decade, the baryon acoustic oscillation (BAO) peak at
about 105{\hMpc} has provided a test of the
Friedmann--Lema\^{\i}tre--Robertson--Walker (FLRW) family of
cosmological models---according to which comoving space is
rigid---strongly favouring a present-day matter density parameter and
dark energy parameter of $\Ommzero\approx0.3$ and
$\OmLamzero\approx0.7$, 
\neededonlyforarXiv{respectively \cite{Eisenstein05,Cole05BAO}.}
{respectively.\cite{Eisenstein05,Cole05BAO}}
However, general relativity does not require comoving space to be
rigid. On the contrary, during the virialisation epoch during which
the most massive structures form by gravitational collapse, it should
be expected that comoving space evolves inhomogeneous curvature as
structure grows inhomogeneously: overdensities contract while underdensities
expand. By averaging over a spatial
\neededonlyforarXiv{slice \cite{Buch00scalav,Buch01scalav},}
{slice,\cite{Buch00scalav,Buch01scalav}}
a generalised Friedmann
equation (Hamiltonian constraint) is found to replace the homogeneous Friedmann
\neededonlyforarXiv{equation \cite{BuchCarf02}.}
{equation.\cite{BuchCarf02}}
Since voids dominate the recent volume,
the effective (averaged) curvature at the present should be
\neededonlyforarXiv{negative \cite{BuchCarf08}.}
{negative.\cite{BuchCarf08}}

The coincidence argument---why does dark energy suddenly become
non-negligible compared to the critical density during the epoch of
galaxy formation?---has been quantified using the virialisation mass
fraction, $\fvir(z)$, of massive dark matter haloes.  This evolves
with decreasing redshift $z$ 
similarly to the dark energy parameter $\OmLam(z)$ interpreted under
FLRW, from a tiny value to a big fraction of unity at the
\neededonlyforarXiv{present \cite{ROB13}.}
{present.\cite{ROB13}}
The Virialisation Approximation, which gives one
example of implementing virialisation in scalar averaging by using the
observed Hubble constant and the peculiar expansion rate of voids as
observational inputs, approximately agrees with the supernovae type Ia
distance-modulus--redshift relation and the present-day effective
matter density 
\neededonlyforarXiv{parameter \cite{ROB13}.}
{parameter.\cite{ROB13}}
Given initial results of other implementations of scalar averaging or Swiss cheese models:
a power-law template 
\neededonlyforarXiv{metric \cite{Larena09template},}
{metric,\cite{Larena09template}}
the Timescape model {\cite{Wiltshire09timescape,DuleyWilt13,NazerW15CMB}}
and the Tardis \cite{LRasSzybka13} model, Occam's razor 
favours dark energy as a phenomenological fit that physically 
represents the recent emergence of average negative curvature.

  \begin{figure}[ht]
    \begin{center}
\begingroup
  \makeatletter
  \providecommand\color[2][]{%
    \GenericError{(gnuplot) \space\space\space\@spaces}{%
      Package color not loaded in conjunction with
      terminal option `colourtext'%
    }{See the gnuplot documentation for explanation.%
    }{Either use 'blacktext' in gnuplot or load the package
      color.sty in LaTeX.}%
    \renewcommand\color[2][]{}%
  }%
  \providecommand\includegraphics[2][]{%
    \GenericError{(gnuplot) \space\space\space\@spaces}{%
      Package graphicx or graphics not loaded%
    }{See the gnuplot documentation for explanation.%
    }{The gnuplot epslatex terminal needs graphicx.sty or graphics.sty.}%
    \renewcommand\includegraphics[2][]{}%
  }%
  \@ifundefined{ifGPcolor}{%
    \newif\ifGPcolor
    \GPcolorfalse
  }{}%
  \@ifundefined{ifGPblacktext}{%
    \newif\ifGPblacktext
    \GPblacktexttrue
  }{}%
  \let\gplgaddtomacro\g@addto@macro
  \gdef\gplbacktext{}%
  \gdef\gplfronttext{}%
  \makeatother
  \ifGPblacktext
  \else
    \ifGPcolor
      \expandafter\def\csname LTw\endcsname{\color{white}}%
      \expandafter\def\csname LTb\endcsname{\color{black}}%
      \expandafter\def\csname LTa\endcsname{\color{black}}%
      \expandafter\def\csname LT0\endcsname{\color[rgb]{1,0,0}}%
      \expandafter\def\csname LT1\endcsname{\color[rgb]{0,1,0}}%
      \expandafter\def\csname LT2\endcsname{\color[rgb]{0,0,1}}%
      \expandafter\def\csname LT3\endcsname{\color[rgb]{1,0,1}}%
      \expandafter\def\csname LT4\endcsname{\color[rgb]{0,1,1}}%
      \expandafter\def\csname LT5\endcsname{\color[rgb]{1,1,0}}%
      \expandafter\def\csname LT6\endcsname{\color[rgb]{0,0,0}}%
      \expandafter\def\csname LT7\endcsname{\color[rgb]{1,0.3,0}}%
      \expandafter\def\csname LT8\endcsname{\color[rgb]{0.5,0.5,0.5}}%
    \else
      \expandafter\def\csname LTw\endcsname{\color{white}}%
      \expandafter\def\csname LTb\endcsname{\color{black}}%
      \expandafter\def\csname LTa\endcsname{\color{black}}%
      \expandafter\def\csname LT0\endcsname{\color{black}}%
      \expandafter\def\csname LT1\endcsname{\color{black}}%
      \expandafter\def\csname LT2\endcsname{\color{black}}%
      \expandafter\def\csname LT3\endcsname{\color{black}}%
      \expandafter\def\csname LT4\endcsname{\color{black}}%
      \expandafter\def\csname LT5\endcsname{\color{black}}%
      \expandafter\def\csname LT6\endcsname{\color{black}}%
      \expandafter\def\csname LT7\endcsname{\color{black}}%
      \expandafter\def\csname LT8\endcsname{\color{black}}%
    \fi
  \fi
  \setlength{\unitlength}{0.0250bp}%
  \begin{picture}(11520.00,8640.00)%
    \gplgaddtomacro\gplbacktext{%
      \colorrgb{0.00,0.00,0.00}%
      \put(1376,1681){\makebox(0,0)[r]{\strut{}0}}%
      \colorrgb{0.00,0.00,0.00}%
      \put(1376,2996){\makebox(0,0)[r]{\strut{}0.02}}%
      \colorrgb{0.00,0.00,0.00}%
      \put(1376,4311){\makebox(0,0)[r]{\strut{}0.04}}%
      \colorrgb{0.00,0.00,0.00}%
      \put(1376,5626){\makebox(0,0)[r]{\strut{}0.06}}%
      \colorrgb{0.00,0.00,0.00}%
      \put(1376,6940){\makebox(0,0)[r]{\strut{}0.08}}%
      \colorrgb{0.00,0.00,0.00}%
      \put(1376,8255){\makebox(0,0)[r]{\strut{}0.1}}%
      \colorrgb{0.00,0.00,0.00}%
      \put(1568,704){\makebox(0,0){\strut{}40}}%
      \colorrgb{0.00,0.00,0.00}%
      \put(2907,704){\makebox(0,0){\strut{}60}}%
      \colorrgb{0.00,0.00,0.00}%
      \put(4247,704){\makebox(0,0){\strut{}80}}%
      \colorrgb{0.00,0.00,0.00}%
      \put(5586,704){\makebox(0,0){\strut{}100}}%
      \colorrgb{0.00,0.00,0.00}%
      \put(6925,704){\makebox(0,0){\strut{}120}}%
      \colorrgb{0.00,0.00,0.00}%
      \put(8264,704){\makebox(0,0){\strut{}140}}%
      \colorrgb{0.00,0.00,0.00}%
      \put(9604,704){\makebox(0,0){\strut{}160}}%
      \colorrgb{0.00,0.00,0.00}%
      \put(10943,704){\makebox(0,0){\strut{}180}}%
      \colorrgb{0.00,0.00,0.00}%
      \put(256,4639){\rotatebox{90}{\makebox(0,0){background-subtracted correlation $\xi - \xi_3$}}}%
      \colorrgb{0.00,0.00,0.00}%
      \put(6255,224){\makebox(0,0){separation $s$ ({\hMpc})}}%
    }%
    \gplgaddtomacro\gplfronttext{%
      \colorrgb{0.00,0.00,0.00}%
      \put(10751,8032){\makebox(0,0)[r]{\strut{}{\footnotesize supercluster-overlap pairs}\ }}%
      \colorrgb{0.00,0.00,0.00}%
      \put(2572,7098){\makebox(0,0){\strut{}A}}%
    }%
    \gplbacktext
    \put(0,0){\includegraphics[scale=0.5]{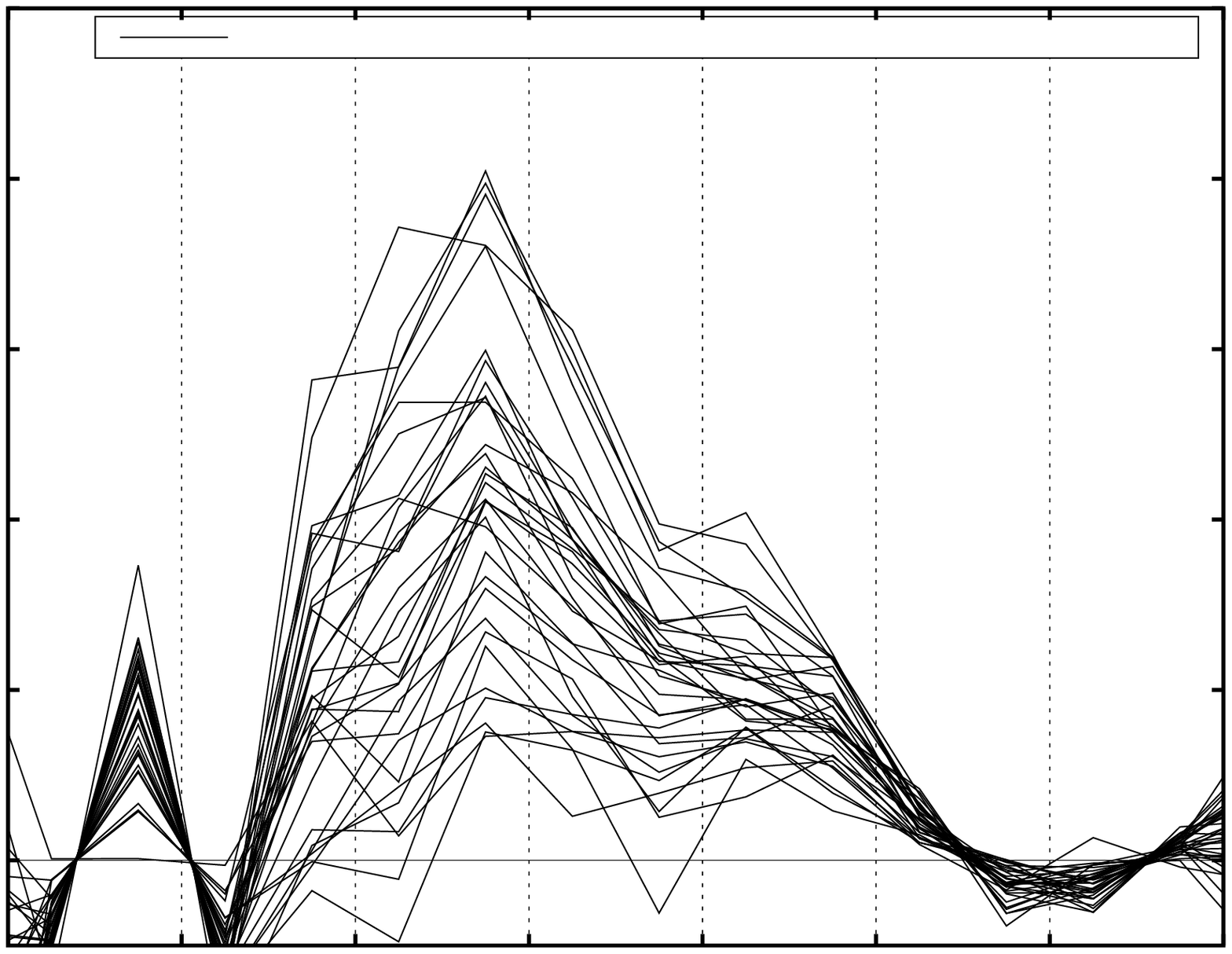}}%
    \gplfronttext
  \end{picture}%
\endgroup

\begingroup
  \makeatletter
  \providecommand\color[2][]{%
    \GenericError{(gnuplot) \space\space\space\@spaces}{%
      Package color not loaded in conjunction with
      terminal option `colourtext'%
    }{See the gnuplot documentation for explanation.%
    }{Either use 'blacktext' in gnuplot or load the package
      color.sty in LaTeX.}%
    \renewcommand\color[2][]{}%
  }%
  \providecommand\includegraphics[2][]{%
    \GenericError{(gnuplot) \space\space\space\@spaces}{%
      Package graphicx or graphics not loaded%
    }{See the gnuplot documentation for explanation.%
    }{The gnuplot epslatex terminal needs graphicx.sty or graphics.sty.}%
    \renewcommand\includegraphics[2][]{}%
  }%
  \@ifundefined{ifGPcolor}{%
    \newif\ifGPcolor
    \GPcolorfalse
  }{}%
  \@ifundefined{ifGPblacktext}{%
    \newif\ifGPblacktext
    \GPblacktexttrue
  }{}%
  \let\gplgaddtomacro\g@addto@macro
  \gdef\gplbacktext{}%
  \gdef\gplfronttext{}%
  \makeatother
  \ifGPblacktext
  \else
    \ifGPcolor
      \expandafter\def\csname LTw\endcsname{\color{white}}%
      \expandafter\def\csname LTb\endcsname{\color{black}}%
      \expandafter\def\csname LTa\endcsname{\color{black}}%
      \expandafter\def\csname LT0\endcsname{\color[rgb]{1,0,0}}%
      \expandafter\def\csname LT1\endcsname{\color[rgb]{0,1,0}}%
      \expandafter\def\csname LT2\endcsname{\color[rgb]{0,0,1}}%
      \expandafter\def\csname LT3\endcsname{\color[rgb]{1,0,1}}%
      \expandafter\def\csname LT4\endcsname{\color[rgb]{0,1,1}}%
      \expandafter\def\csname LT5\endcsname{\color[rgb]{1,1,0}}%
      \expandafter\def\csname LT6\endcsname{\color[rgb]{0,0,0}}%
      \expandafter\def\csname LT7\endcsname{\color[rgb]{1,0.3,0}}%
      \expandafter\def\csname LT8\endcsname{\color[rgb]{0.5,0.5,0.5}}%
    \else
      \expandafter\def\csname LTw\endcsname{\color{white}}%
      \expandafter\def\csname LTb\endcsname{\color{black}}%
      \expandafter\def\csname LTa\endcsname{\color{black}}%
      \expandafter\def\csname LT0\endcsname{\color{black}}%
      \expandafter\def\csname LT1\endcsname{\color{black}}%
      \expandafter\def\csname LT2\endcsname{\color{black}}%
      \expandafter\def\csname LT3\endcsname{\color{black}}%
      \expandafter\def\csname LT4\endcsname{\color{black}}%
      \expandafter\def\csname LT5\endcsname{\color{black}}%
      \expandafter\def\csname LT6\endcsname{\color{black}}%
      \expandafter\def\csname LT7\endcsname{\color{black}}%
      \expandafter\def\csname LT8\endcsname{\color{black}}%
    \fi
  \fi
  \setlength{\unitlength}{0.0250bp}%
  \begin{picture}(11520.00,8640.00)%
    \gplgaddtomacro\gplbacktext{%
      \colorrgb{0.00,0.00,0.00}%
      \put(1376,1681){\makebox(0,0)[r]{\strut{}0}}%
      \colorrgb{0.00,0.00,0.00}%
      \put(1376,2996){\makebox(0,0)[r]{\strut{}0.02}}%
      \colorrgb{0.00,0.00,0.00}%
      \put(1376,4311){\makebox(0,0)[r]{\strut{}0.04}}%
      \colorrgb{0.00,0.00,0.00}%
      \put(1376,5626){\makebox(0,0)[r]{\strut{}0.06}}%
      \colorrgb{0.00,0.00,0.00}%
      \put(1376,6940){\makebox(0,0)[r]{\strut{}0.08}}%
      \colorrgb{0.00,0.00,0.00}%
      \put(1376,8255){\makebox(0,0)[r]{\strut{}0.1}}%
      \colorrgb{0.00,0.00,0.00}%
      \put(1568,704){\makebox(0,0){\strut{}40}}%
      \colorrgb{0.00,0.00,0.00}%
      \put(2907,704){\makebox(0,0){\strut{}60}}%
      \colorrgb{0.00,0.00,0.00}%
      \put(4247,704){\makebox(0,0){\strut{}80}}%
      \colorrgb{0.00,0.00,0.00}%
      \put(5586,704){\makebox(0,0){\strut{}100}}%
      \colorrgb{0.00,0.00,0.00}%
      \put(6925,704){\makebox(0,0){\strut{}120}}%
      \colorrgb{0.00,0.00,0.00}%
      \put(8264,704){\makebox(0,0){\strut{}140}}%
      \colorrgb{0.00,0.00,0.00}%
      \put(9604,704){\makebox(0,0){\strut{}160}}%
      \colorrgb{0.00,0.00,0.00}%
      \put(10943,704){\makebox(0,0){\strut{}180}}%
      \colorrgb{0.00,0.00,0.00}%
      \put(256,4639){\rotatebox{90}{\makebox(0,0){background-subtracted correlation $\xi - \xi_3$}}}%
      \colorrgb{0.00,0.00,0.00}%
      \put(6255,224){\makebox(0,0){separation $s$ ({\hMpc})}}%
    }%
    \gplgaddtomacro\gplfronttext{%
      \colorrgb{0.00,0.00,0.00}%
      \put(10751,8032){\makebox(0,0)[r]{\strut{}{\footnotesize complementary pairs}\ }}%
      \colorrgb{0.00,0.00,0.00}%
      \put(2572,7098){\makebox(0,0){\strut{}B}}%
    }%
    \gplbacktext
    \put(0,0){\includegraphics[scale=0.5]{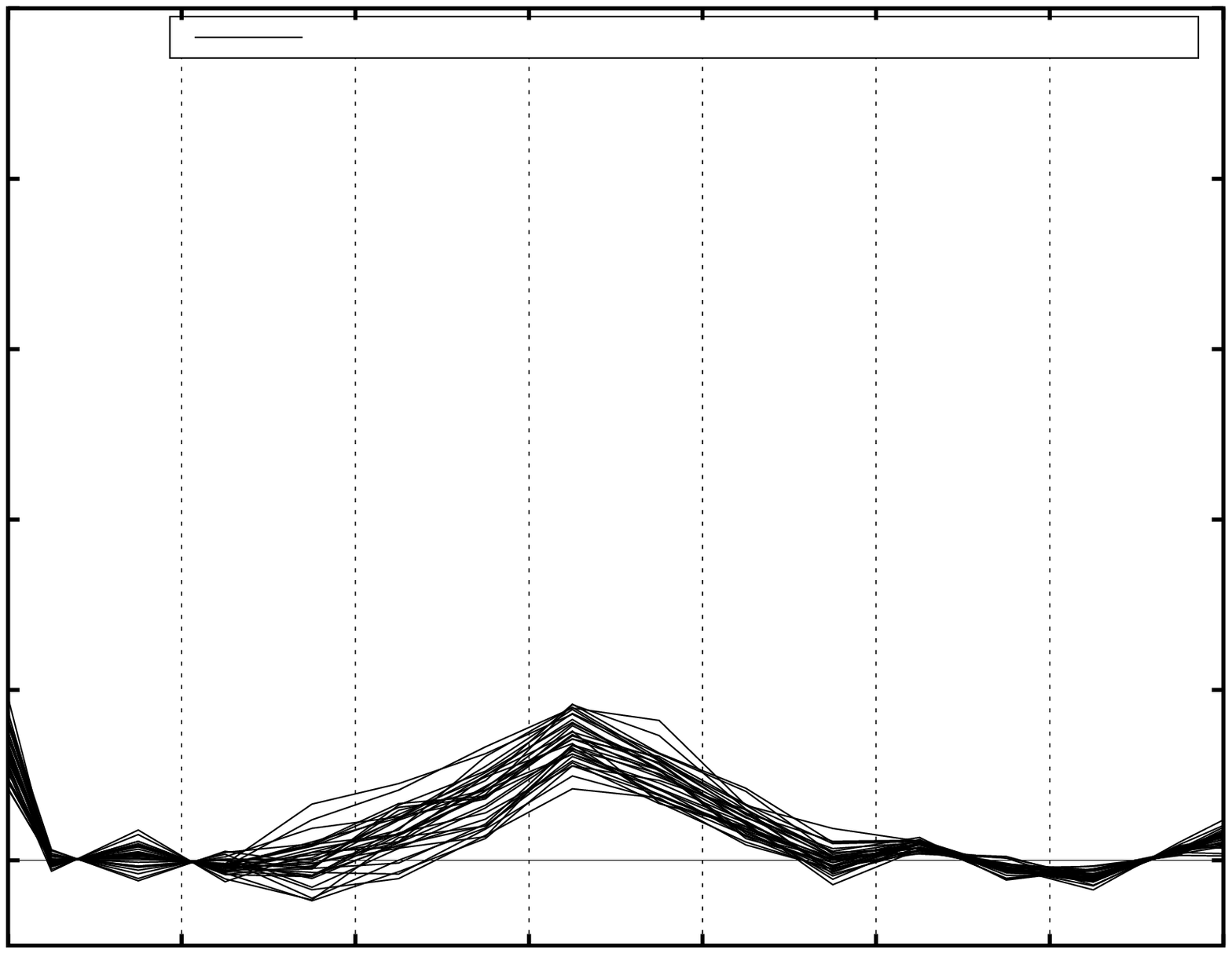}}%
    \gplfronttext
  \end{picture}%
\endgroup
\end{center}
    \caption{BAO peak shift, as in 
      Fig.~1, \protect\inlinecite{RBFO15}.
      A {\em (above)}:
      BAO peak for LRG pairs whose paths' overlap with
      a supercluster is $\omega \ge \omega_{\min} =60${\hMpc} or are
      completely contained in a supercluster, where
      the overlap $\omega$ is defined in
      Sect.~2.3, Fig.~1 of \protect\inlinecite{RBOF15}.
      Individual curves show
      supercluster and ``random'' galaxy
      bootstrap resampling. The BAO peak mostly occurs at 95{\hMpc}, shortward
      of the usual value.
      B {\em (below)}: Complementary LRG pair subset.
      The peak occurs at the usual value of 105{\hMpc}.
      \label{f-thr60}}
\end{figure}

A key order-of-magnitude 
\prerefereechanges{argument---how can the volume-weighted average curvature parameter 
(e.g. (123) in \inlinecite{BuchCarf02}; (2.9) in \inlinecite{ROB13}) grow to a high enough amplitude in comparison to the
matter density parameter?---follows} from the 
similarity in order of magnitude between the Hubble constant 
and the void peculiar expansion rate 
{(see (2.27), (2.22), (2.13a) in \inlinecite{ROB13})}: both are indisputably 
several tens of km/s/Mpc; the former is well-studied, the latter
is poorly studied.
Another promising observational avenue for relativistic cosmology is
measuring the emergence of average negative curvature, i.e. 
using the Clarkson, {Bassett} \& Lu
$\OmegakClarkson(z)$ 
\neededonlyforarXiv{relation \cite{ClarksonBL08,Clarkson12test,SaponeMN14}.}
{relation.\cite{ClarksonBL08,Clarkson12test,SaponeMN14}}
However, it should be possible to use 
a standard ruler to measure the inhomogeneity itself, rather 
\prerefereechanges{than}
average curvature.

\section{BAO peak location inhomogeneity}

\paragraph{Relativistic Zel'dovich approximation lower limit}
The BAO peak standard ruler provides this: a method of detecting
curvature inhomogeneity by measuring the environment dependence of the
\neededonlyforarXiv{scale factor \cite{RBOF15,RBFO15}.}
{scale factor.\cite{RBOF15,RBFO15}} 
Writing $\CM$ (``Massive'') and
$\CE$ {(``Empty'')} to represent overdense and underdense spatial
regions, respectively, the scalar averaged scale factors $a_{\CM}$
and $a_{\CE}$ should be low and high, respectively, i.e.  $a_{\CM} <
a_{\CE}$. Using equations (2), (13), (32), (50), and (54) of
\inlinecite{BuchRZA2} to integrate the Raychaudhuri
equation\footnote{Eq.~(9) of \protect\inlinecite{BuchRZA2}.} and
1$\sigma$ initial overdensities/underdensities in a 105{\hMpc}
diameter spherical domain gives 
a relativistic Zel'dovich approximation \cite{BuchRZA1,BuchRZA2}
estimate of inhomogeneous scale 
\neededonlyforarXiv{factors $a_{\CM} \approx 0.91 a_{\CE}$ \cite{RBFO15}.}
{factors $a_{\CM} \approx 0.91 a_{\CE}$.\cite{RBFO15}}

\paragraph{Method} 
The Sloan Digital Sky Survey (SDSS) Data Release 7 (DR7) Luminous
Red Galaxies (LRGs) provide a well-defined BAO peak. 
The two-point auto-correlation function $\xi$ of 
pairs of 
SDSS DR7 LRGs that are preferentially tangential to the observer's
line-of-sight should have a peak that is less affected by redshift
space distortions than that for radial pairs. Using 
\inlinecite{Kazin2010}'s catalogues of real and random galaxies,
defining overlap $\omega$ between LRG pairs and 
\inlinecite{NadHot2013}'s supercluster catalogue
(Sect.~2.3, Fig.~1, \inlinecite{RBOF15}),
assuming three-dimensional comoving 
separations $s$ for a standard $\Lambda$CDM
model 
\neededonlyforarXiv{($\Ommzero=0.32, \OmLamzero= 0.68$) \citep{WMAPSpergel,PlanckXVIcosmoparam13},}
{($\Ommzero=0.32, \OmLamzero= 0.68$),\citep{WMAPSpergel,PlanckXVIcosmoparam13}}
with the justification
that this is a phenomenologically reasonable fit, 
using the Landy \& Szalay correlation 
\neededonlyforarXiv{estimator \citep{LandySz93},}
{estimator,\citep{LandySz93}}
and subtracting a cubic fit to $\xi$ from ranges of separation $s$ away 
from the peak ($s < 70${\hMpc} and $s > 140${\hMpc}) yields
BAO peaks such as those shown in Fig.~\ref{f-thr60}.

{
\begin{figure}
  \begin{center}
\begingroup
  \makeatletter
  \providecommand\color[2][]{%
    \GenericError{(gnuplot) \space\space\space\@spaces}{%
      Package color not loaded in conjunction with
      terminal option `colourtext'%
    }{See the gnuplot documentation for explanation.%
    }{Either use 'blacktext' in gnuplot or load the package
      color.sty in LaTeX.}%
    \renewcommand\color[2][]{}%
  }%
  \providecommand\includegraphics[2][]{%
    \GenericError{(gnuplot) \space\space\space\@spaces}{%
      Package graphicx or graphics not loaded%
    }{See the gnuplot documentation for explanation.%
    }{The gnuplot epslatex terminal needs graphicx.sty or graphics.sty.}%
    \renewcommand\includegraphics[2][]{}%
  }%
  \@ifundefined{ifGPcolor}{%
    \newif\ifGPcolor
    \GPcolorfalse
  }{}%
  \@ifundefined{ifGPblacktext}{%
    \newif\ifGPblacktext
    \GPblacktexttrue
  }{}%
  \let\gplgaddtomacro\g@addto@macro
  \gdef\gplbacktext{}%
  \gdef\gplfronttext{}%
  \makeatother
  \ifGPblacktext
  \else
    \ifGPcolor
      \expandafter\def\csname LTw\endcsname{\color{white}}%
      \expandafter\def\csname LTb\endcsname{\color{black}}%
      \expandafter\def\csname LTa\endcsname{\color{black}}%
      \expandafter\def\csname LT0\endcsname{\color[rgb]{1,0,0}}%
      \expandafter\def\csname LT1\endcsname{\color[rgb]{0,1,0}}%
      \expandafter\def\csname LT2\endcsname{\color[rgb]{0,0,1}}%
      \expandafter\def\csname LT3\endcsname{\color[rgb]{1,0,1}}%
      \expandafter\def\csname LT4\endcsname{\color[rgb]{0,1,1}}%
      \expandafter\def\csname LT5\endcsname{\color[rgb]{1,1,0}}%
      \expandafter\def\csname LT6\endcsname{\color[rgb]{0,0,0}}%
      \expandafter\def\csname LT7\endcsname{\color[rgb]{1,0.3,0}}%
      \expandafter\def\csname LT8\endcsname{\color[rgb]{0.5,0.5,0.5}}%
    \else
      \expandafter\def\csname LTw\endcsname{\color{white}}%
      \expandafter\def\csname LTb\endcsname{\color{black}}%
      \expandafter\def\csname LTa\endcsname{\color{black}}%
      \expandafter\def\csname LT0\endcsname{\color{black}}%
      \expandafter\def\csname LT1\endcsname{\color{black}}%
      \expandafter\def\csname LT2\endcsname{\color{black}}%
      \expandafter\def\csname LT3\endcsname{\color{black}}%
      \expandafter\def\csname LT4\endcsname{\color{black}}%
      \expandafter\def\csname LT5\endcsname{\color{black}}%
      \expandafter\def\csname LT6\endcsname{\color{black}}%
      \expandafter\def\csname LT7\endcsname{\color{black}}%
      \expandafter\def\csname LT8\endcsname{\color{black}}%
    \fi
  \fi
  \setlength{\unitlength}{0.0250bp}%
  \begin{picture}(11520.00,8640.00)%
    \gplgaddtomacro\gplbacktext{%
      \colorrgb{0.00,0.00,0.00}%
      \put(992,1024){\makebox(0,0)[r]{\strut{}0}}%
      \colorrgb{0.00,0.00,0.00}%
      \put(992,2832){\makebox(0,0)[r]{\strut{}5}}%
      \colorrgb{0.00,0.00,0.00}%
      \put(992,4640){\makebox(0,0)[r]{\strut{}10}}%
      \colorrgb{0.00,0.00,0.00}%
      \put(992,6447){\makebox(0,0)[r]{\strut{}15}}%
      \colorrgb{0.00,0.00,0.00}%
      \put(992,8255){\makebox(0,0)[r]{\strut{}20}}%
      \colorrgb{0.00,0.00,0.00}%
      \put(1184,704){\makebox(0,0){\strut{}0}}%
      \colorrgb{0.00,0.00,0.00}%
      \put(2958,704){\makebox(0,0){\strut{}20}}%
      \colorrgb{0.00,0.00,0.00}%
      \put(4733,704){\makebox(0,0){\strut{}40}}%
      \colorrgb{0.00,0.00,0.00}%
      \put(6507,704){\makebox(0,0){\strut{}60}}%
      \colorrgb{0.00,0.00,0.00}%
      \put(8281,704){\makebox(0,0){\strut{}80}}%
      \colorrgb{0.00,0.00,0.00}%
      \put(10056,704){\makebox(0,0){\strut{}100}}%
      \colorrgb{0.00,0.00,0.00}%
      \put(256,4639){\rotatebox{90}{\makebox(0,0){peak shift ${\Delta}s$ ({\hMpc})}}}%
      \colorrgb{0.00,0.00,0.00}%
      \put(6063,224){\makebox(0,0){minimum overlap $\omega_{\mathrm{min}}$ ({\hMpc})}}%
    }%
    \gplgaddtomacro\gplfronttext{%
      \colorrgb{0.00,0.00,0.00}%
      \put(10751,8032){\makebox(0,0)[r]{\strut{}{\footnotesize fit to SDSS DR7 LRGs}\ }}%
      \colorrgb{0.00,0.00,0.00}%
      \put(10751,7712){\makebox(0,0)[r]{\strut{}{\footnotesize RZA theor.\ lower limit}\ }}%
    }%
    \gplbacktext
    \put(0,0){\includegraphics[scale=0.5]{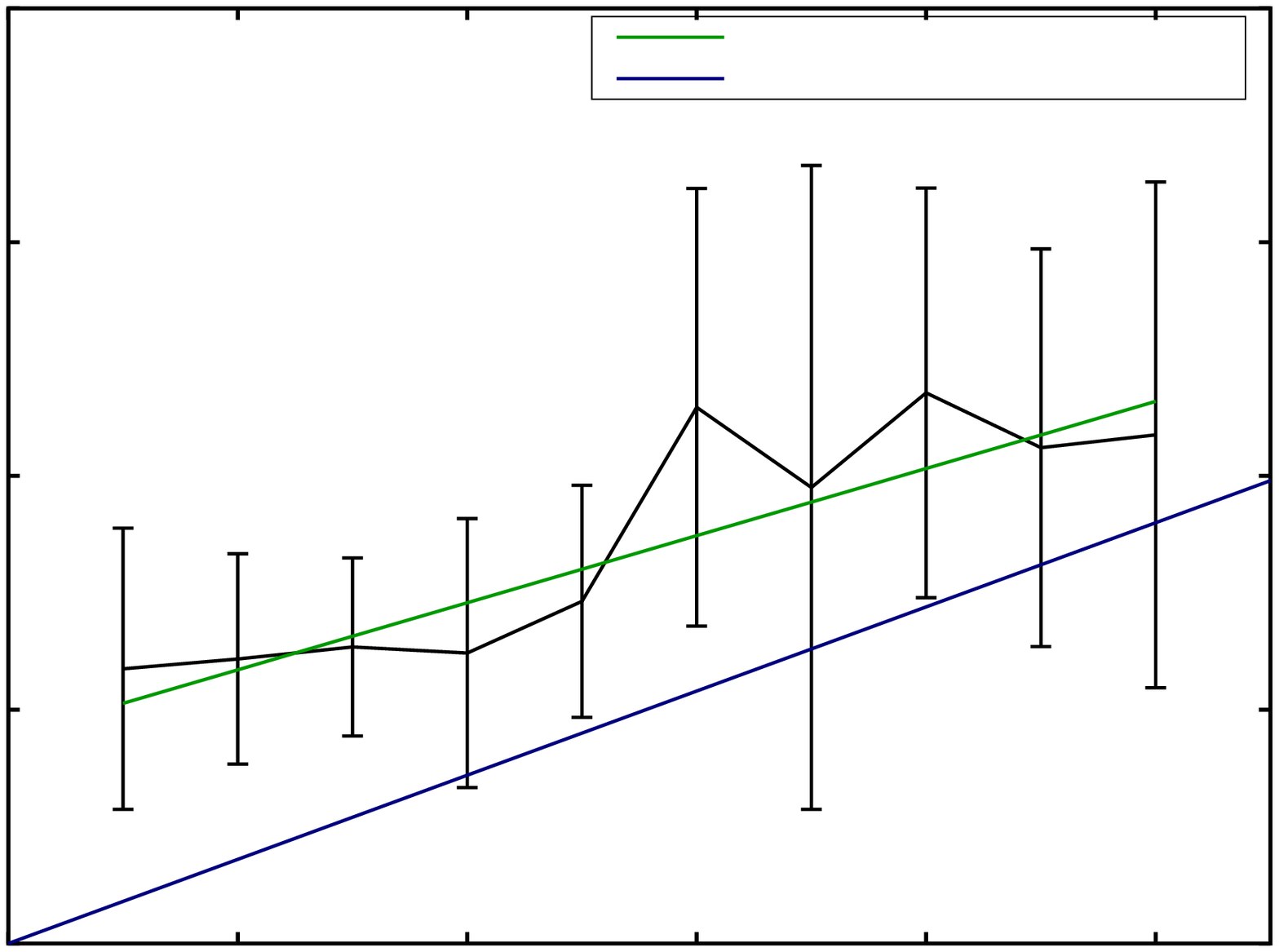}}%
    \gplfronttext
  \end{picture}%
\endgroup
  \end{center}
  \caption{Environment dependence of the BAO peak
    shift, as in Fig.~2, \protect\inlinecite{RBFO15}.
    The BAO peak shift $\Delta s :=
    s\textsubscript{non-sc} - s_{\mathrm{sc}}$, where
    $s_{\mathrm{sc}}$ and $s\textsubscript{non-sc}$ are
    estimated with best-fit Gaussians for LRG pairs overlapping
    (sc) or not overlapping (non-sc) superclusters, with
    robust estimates of the standard errors.
    A best-fit line {$\Delta s =
      4.3${\hMpc}$+ 0.07 \omega_{\min}$} is shown in green.
    A 9\% shift (i.e. $a_{\CM}/a_{\CE} = 0.91$) 
    would give
      $\Delta s = 0.09 \omega$, i.e. 
      a lower limit from scalar-averaging, shown in blue.
    \label{f-overlap-dependence}}
\end{figure}
} 

\paragraph{Results}
In Fig.~\ref{f-thr60}, 
the BAO peak location clearly shifts to smaller values
for environments which, according to scalar averaging, should
have lower values of the effective scale factor than that 
in the effective model, i.e. we see that $a_{\CM} < a_{\CE}$,
and, as expected in a void-dominated model,
$a_{\CE} \approx a_{\Lambda\mathrm{CDM}}$. The reality of this
shift can be tested further by considering its dependence
on the minimum overlap required for considering an LRG pair
to overlap with a supercluster.

Figure~\ref{f-overlap-dependence} shows this dependence.
The more that LRG pairs overlap with superclusters, the lower
the averaged scale factor, in contrast with the rigid comoving
space assumption of the FLRW model. The statistical significance
of this relation can be estimated with the Pearson
product--moment correlation coefficient for $\Delta s$ and
$\omega_{\min}$, estimated as 0.87. The null hypothesis that
there is no correlation has a probability of $P \approx
0.0008$, i.e. the detection is highly significant.
The linear least-squares best fit has slope and zero point 
$0.073 \pm 0.040$ and $4.3 \pm 2.0${\hMpc}, respectively.
The BAO peak standard ruler is flexible.

\section{Conclusion}
Work towards a relativistically more accurate cosmological model than
$\Lambda$CDM is still in progress, but the early results are promising.
This initial detection of inhomogeneity in the scale factor should, during
the coming decades---with 
\neededonlyforarXiv{Euclid \cite{EuclidScienceBook2010},}
{Euclid,\cite{EuclidScienceBook2010}}
\neededonlyforarXiv{eBOSS (extended Baryon Oscillation Spectroscopic Survey) \cite{Zhao15eBOSSpredict},
DESI (Dark Energy Spectroscopic Instrument) \cite{Levi13DESI},
4MOST (4-metre Multi-Object Spectroscopic Telescope) \cite{deJong12VISTA4MOST,deJong14subscriponly4MOST},}
{eBOSS (extended Baryon Oscillation Spectroscopic Survey),\cite{Zhao15eBOSSpredict}
DESI (Dark Energy Spectroscopic Instrument),\cite{Levi13DESI}
4MOST (4-metre Multi-Object Spectroscopic Telescope),\cite{deJong12VISTA4MOST,deJong14subscriponly4MOST},}
and the LSST (Large Synoptic Survey Telescope) \cite{TysonLSST03}---be 
able to help distinguish which implementations 
best match observations, as well as increase the statistical confidence in 
rejecting \cite{BCKRW16,Bull15BeyondLCDM}
the 
{\em {Newtonian-structure-formation decoupled
from}
relativistic-expansion\footnote{{Scalar 
    averaging implies that these are 
    \neededonlyforarXiv{coupled \cite{Buch01scalav,RoyBuch12arbbackground}.}
                       {coupled.\cite{Buch01scalav,RoyBuch12arbbackground}}}}
hypothesis} 
{fundamental to} $\Lambda$CDM. 
{Although models} 
that speculate beyond both the Newtonian and general-relativistic models are 
presently very 
{popular, the prospects for a dark-energy--free
general-relativistic cosmological model look good.}

\section*{Acknowledgments}

BFR acknowledges support for a part of this project from the
HECOLS International Associated Laboratory, supported in part by the
National Science Centre, Poland, grant DEC--2013/08/M/ST9/00664, and
for a part of the project under grant 2014/13/B/ST9/00845 of the
National Science Centre, Poland, and visiting 
support from the Centre de Recherche Astrophysique de Lyon.
{Thanks to Thomas Buchert, Jan Ostrowski and
David Wiltshire for useful comments.}

\end{document}